# Effects of transition-metal spacers on the spin-orbit torques, spin Hall magnetoresistance, and magnetic anisotropy of Pt/Co bilayers

Can Onur Avci[1,2], Geoffrey S. D. Beach[1] and Pietro Gambardella[2]

[1]Department of Materials Science and Engineering, Massachusetts Institute of Technology, Cambridge, Massachusetts 02139, USA

[2] Department of Materials, ETH Zürich, CH-8093 Zürich, Switzerland

**We studied the effect of inserting 0.5 nm-thick spacer layers (Ti, V, Cr, Mo, W) at the Pt/Co interface on the spin-orbit torques, Hall effect, magnetoresistance, saturation magnetization, and magnetic anisotropy. We find that the damping-like spin-orbit torque decreases substantially for all samples with a spacer layer compared to the reference Pt/Co bilayer, consistently with the opposite sign of the atomic spin-orbit coupling constant of the spacer elements relative to Pt. The reduction of the damping-like torque is monotonic with atomic number for the isoelectronic 3*d*, 4*d*, and 5*d* elements, with the exception of V that has a stronger effect than Cr. The field-like spin-orbit torque almost vanishes for all spacer layers irrespective of their composition, suggesting that this torque predominantly originates at the Pt/Co interface. The anomalous Hall effect, magnetoresistance, and saturation magnetization are also all reduced substantially, whereas the sheet resistance is increased in the presence of the spacer layer. Finally, we evidence a correlation between the amplitude of the spin-orbit torques, the spin Hall-like magnetoresistance, and the perpendicular magnetic anisotropy. These results highlight the significant influence of ultrathin spacer layers on the magnetotransport properties of heavy metal/ferromagnetic systems.**

# I - INTRODUCTION

Current-induced spin-orbit torques (SOTs) have emerged as a powerful tool to manipulate the magnetization of heavy metal/ferromagnet (HM/FM) bilayers characterized by strong spin-orbit coupling and structural inversion asymmetry[1–8]. Interfaces play a crucial role in determining the strength and symmetries of SOTs[7,9–11], as well as other interface-related spin transport and dynamic effects such as the spin Hall[12] and Rashba-Edelstein magnetoresistance[13], unidirectional magnetoresistance[14–20], spin Seebeck effect[21], spin-torque ferromagnetic resonance[22], and spin pumping[23–25]. Additionally, interfaces in thin film structures play a dominant role in many other magnetic and electrical properties such as perpendicular magnetic anisotropy[26–30], proximity magnetism[30–33], anisotropic magnetoresistance[34–37], and anomalous Hall effect[38–43].

The damping-like (DL) and field-like (FL) SOT are manifestations of the spin accumulation generated by an in-plane charge current flowing through HM/FM bilayers[9,44–46]. The most widely used HM layers are $5d$ elements such as Pt, Ta or W (Refs. [2,4,6,7,45,47–51]), although, more recently, lighter metals such as V, Cr, Mo, and Pd have also been shown to generate substantial SOTs[52–55]. The SOTs in HM/FM heterostructures originate from the spin Hall effect (SHE) in the bulk of the HM and from interfacial spin currents arising from spin-dependent scattering and Rashba-type spin-orbit coupling due to broken structural inversion symmetry[7,56–61]. All such effects generate a spin accumulation at the HM/FM interface that contributes to both types of torques[62]. Independent of their origins, SOTs are highly interface-sensitive since the spin accumulation occurs at or near the interface.

Spacer layers in HM/FM systems have been widely used in order to minimize magnetic proximity effects[63,64] and/or separate the HM as a source of spin current from the FM[10,48,65]. In most such cases, Cu has been the spacer element of choice owing to its weak induced magnetic moment[66] and long spin diffusion length[67]. Other elements employed as spacers are Hf (Ref. [68]) and Au (Ref. [10]), which have been shown to improve the magnitude of the SOT in Pt/Hf/CoFeB and Pt/Au/Co/Ni/Co, respectively. Whereas the latter results have been interpreted in terms of an increase of the spin transparency of the interfaces within a drift-diffusion formalism[10,11,48,69–71], recent theoretical and experimental studies point out that the presence of spin-orbit coupling additionally leads to the rotation, flipping, and generation of spins at interfaces[61,72–76]. As interfacial spin-orbit coupling plays a role in many different phenomena apart from SOT, such as magnetoresistance, anomalous Hall effect, and magnetic anisotropy, investigations of spacer layers provide insight into the correlation of such effects while offering alternative ways to control the interfacial spin transport properties in HM/FM bilayers.

In this paper, we present a systematic investigation of the influence of ultrathin spacer layers on the SOTs, magnetoresistance, Hall effect, saturation magnetization, and magnetic anisotropy of the archetypal Pt/Co bilayer system. We used five different spacer elements (Ti, V, Cr, Mo, W), of which the first three are non-magnetic $3d$ elements with increasing atomic number and orbital filling, whereas the last three are isoelectronic group-IV elements with $3d$, $4d$, $5d$ valence. We find that the DL-SOT depends strongly on the choice of spacer layer, decreasing monotonically from the $3d$ to $5d$ elements, but with no clear dependence on the atomic number within the $3d$ series. In contrast to the DL-SOT, the FL-SOT becomes negligibly small, independently of the type of spacer layer, indicating that it predominantly originates at the Pt/Co interface. We also measure a large magnetoresistance upon rotating the magnetization in the plane perpendicular to the current, which is typically associated with the spin Hall magnetoresistance (SMR). We reveal a clear correlation between this magnetoresistance and the DL-SOT, showing that the current-induced spin accumulation plays an important role in this phenomenon. Further analysis shows that the SMR alone cannot be responsible for this unconventional magnetoresistance. Rather, our results show that interface contributions play a significant role over the SMR originating from the bulk SHE. Finally, we reveal that the DL-SOT is also correlated with the interfacial perpendicular magnetic anisotropy, evidencing that the spin torque generation at the Pt/Co interface may be related with the same interfacial spin-orbit coupling mechanism giving rise to the perpendicular anisotropy. These findings highlight the importance of the interfaces in spin transport and magneto-electric properties in HM/FM bilayer systems and provide insight into controlling the above properties by interface engineering.

This paper is organized as follows. Section II presents the experimental details concerning the layer growth, device preparation and measurement procedures. Section III.A reports the magnetic and electrical characterization of the layers by means of vibrating sample magnetometry, Hall effect, and resistivity measurements. Sections III:B-D present the magnetoresistance and SOT measurements, their analysis, and a discussion as to how the different properties correlate. Concluding remarks are given in Sect. IV.

## II – EXPERIMENTAL DETAILS AND METHODS

We grew //Ta(2)/Pt(6)/$X$(0.5)/Co(2)/Ta(2.5) multilayers on thermally oxidized Si wafers by dc magnetron sputtering (Fig.1 – right panel). Here, the numbers in brackets are thicknesses in nm and $X$ denotes the element used as the spacer layer, i.e., $X$ = Ti, V, Cr, Mo, W. The Ta under and over layers serve as buffer and capping, respectively. The sputtering chamber base pressure was 2.5-4×$10^{-7}$ mbar and the Ar partial pressure was 4×$10^{-3}$ mbar. The deposition rate was ~2 nm/min and the applied power was ~150 W for $X$, 23 W for Pt, and ~116 W for Co. The target to substrate distance for Pt was about 10 cm, whereas for Co and $X$ it was ~20 cm. For each stack, we simultaneously prepared a second structure that does not include the spacer layer, by masking one of the samples during the deposition of element $X$. Thus, the influence of the latter can be accurately examined by comparing the properties of each sample pair, with and without spacer, prepared in identical conditions. We note that the sputter deposition method used here can lead to partial intermixing of the neighboring layers[77]. While it is hard to have an exact quantitative measure of intermixing in ultrathin systems, previous studies evidenced that for instance Co (or other similar FM) deposition on Ti (ref.[78,79]), V (ref.[80,81]), Cr (ref.[82,83]), Mo (ref.[84,85]), W (ref.[86,87]) result in interfacial mixing on the order of 0.5 nm, whereas the mixing of Co and Pt is usually limited to the topmost surface layer[88]. Likewise, although literature studies are scarcer, the deposition of these spacer layers on Pt can lead to intermixing. Based on these studies, we assume that the insertion of $X$ cannot be strictly treated as an additional layer, but should be rather considered as a transition region between Pt and Co with a rich content of $X$ near the interface.

The Hall bar structures, shown in Fig.1 (a), were fabricated using standard optical lithography and lift-off with the current line width $w = 50$ μm and distance between the two Hall arms $l = 250$ μm. Simultaneously, we also grew continuous films to measure the saturation magnetization ($M_s$) of each layer. All samples have easy-plane magnetic anisotropy as the thickness of Co is larger than the threshold (~1 nm) of the out-of-plane to in-plane spin reorientation transition of Pt/Co. For the electrical measurements, the Hall bars were wire bonded and mounted on a motorized stage allowing for in-plane ($\varphi$) and out-of-plane ($\theta$) rotation, and placed in an electromagnet producing fields of up to 2 T. Figure 1 (a) shows the definition of the angles and coordinate system. Experiments were performed at room temperature using an ac current density of amplitude $j = 2.7 - 2.9 \times 10^6$ A/cm² and frequency $\omega/2\pi = 10$ Hz. In the following, the current density is obtained by dividing the total current by the cross section of the Pt, spacer, and Co layers. Current shunting by the buffer and cap layers is neglected due the high resistivity of Ta and their partial oxidation through $SiO_2$ reduction at the substrate interface (bottom Ta) and exposure to atmosphere (top Ta).

In order to characterize the magnetotransport properties of Pt/Co and Pt/$X$/Co, we recorded the first and second harmonic Hall resistances ($R_{\omega,H}$, $R_{2\omega,H}$) and the first harmonic longitudinal resistance ($R_{\omega,L}$). The first harmonic Hall resistance consists of the anomalous Hall ($R_{AHE}$) and planar Hall effect ($R_{PHE}$) contributions and is defined as follows:

$$R_{\omega,H} = R_{AHE} \cos\theta + R_{PHE} \sin^2\theta \sin 2\varphi. \tag{1}$$

The second harmonic Hall resistance reflects the SOT-induced oscillations of the magnetization as well as the magneto-thermal voltage due to the thermal gradients induced by Joule heating. This term depends explicitly on the damping-like and field-like SOT effective fields (the latter including the Oersted field due to current flow in the nonmagnetic layers), and the magneto-thermal effects, predominantly driven by an out-of-plane temperature gradient ($\nabla T_z$)[89]:

$$R_{2\omega,H} = \left[\left(R_{AHE}\frac{b_{DL}}{B_{eff}} + \frac{\alpha \nabla T_z}{I_0}\right)\cos\varphi + 2R_{PHE}\frac{b_{FL}}{B_{ext}}(2\cos^3\varphi - \cos\varphi)\right]. \tag{2}$$

Here, $b_{DL}$ and $b_{FL}$ are the ratios of the damping-like and field-like SOT effective fields to the applied current, respectively, and $\alpha$ is the magneto-thermal coefficient taking into account the anomalous Nernst and spin Seebeck effects. These two effects are considered together as they share the same angular dependence and cannot be easily distinguished in our measurements. $B_{eff}$ is the effective static fields acting on the magnetization and is the sum of the external field, demagnetizing field and anisotropy fields: $B_{eff} = B_{ext} + B_{dem} + B_{ani}$. We assume that Joule heating by the injected current is the only source of temperature gradient, hence $\nabla T_z \propto j^2 R$, where $R$ is the device resistance. We note that Eq. 2 is valid when the magnetization lies in the *xy*-plane. In such a case, the most convenient way to separate the SOT and magneto-thermal contributions is to perform *xy* angular scan measurements with a rotating field $B_{ext}$ of fixed amplitude. We show and discuss the representative $R_{\omega,H}$ and $R_{2\omega,H}$ data in Sec. IV B. A more detailed description of the analysis and quantification of SOTs and magneto-thermal effects is reported elsewhere[89].

The first harmonic longitudinal resistance is equivalent to the standard dc measurement and can be written in its most general form as[14]:

$$R_{\omega,L} = R_0 - \Delta R_{zx}\sin^2\theta\cos^2\varphi - \Delta R_{zy}\sin^2\theta\sin^2\varphi, \tag{3}$$

where $R_0 \equiv R(\mathbf{m}//\mathbf{x})$, $\Delta R_{zx}$ is the resistance difference between magnetization pointing along the *z*-axis and the *x*-axis, and similarly, $\Delta R_{zy}$ is the resistance difference between magnetization pointing along the *z*-axis and the *y*-axis. We note that the straightforward derivation of $\Delta R_{xy}$ can be made by simply subtracting $\Delta R_{zx}$ from $\Delta R_{zy}$, such as $\Delta R_{xy} = \Delta R_{zy} - \Delta R_{zx}$.

### III – RESULTS AND DISCUSSION

A. **Magnetic and electrical properties**

We determined the saturation magnetization ($M_s$) of each layer by measuring in-plane hysteresis loops using a vibrating sample magnetometer. Figure 2 (a) shows exemplary hysteresis loops for Pt/Ti/Co (red squares) and Pt/Co (black circles). Figure 2 (b) shows the $M_s$ of all the samples studied in this work. For each element, we plot the two values corresponding to the samples with (red squares) and without (black circles) spacer layer measured on the sample pairs deposited at the same time. Note that we adapt this data presentation style in the remainder of the paper, when applicable. With the exception of the W sample pair, we measure a larger $M_s$ for all the samples without spacer layer; on average, we estimate $M_s$[Pt/Co] ~ 1.3×10$^6$ A/m and $M_s$[Pt/$X$/Co] ~ 1.1×10$^6$ A/m. We associate the different $M_s$ between the samples with and without the spacer to the induced moment in Pt when it is in direct contact with Co (Refs. [30–33]). The difference in $M_s$ (~0.2×10$^6$ A/m) corresponds to about 0.64 $\mu_B$ per Pt atom, assuming that 1 nm of Pt is magnetized, which is in close agreement with literature values[90]. Notwithstanding the induced magnetization in Pt, the average $M_s$ of Pt/Co is about 10% smaller compared to bulk Co. We attribute this reduction to the presence of a magnetic dead layer at the interface between Co and the Ta capping layer and Co/Ta intermixing, as shown for previous studies of Pt/Co/Ta (Refs. [8,29]). For certain elements, it is also possible that the magnetic moments of the Co atoms in contact with the spacer layer are reduced in comparison with their bulk values[29]. This effect may also contribute to the reduced $M_s$ of the Pt/$X$/Co samples, together with intermixing.

We next measure the anomalous Hall resistance ($R_{AHE}$) by sweeping the out-of-plane field ($B_z$). Figure 2 (c) shows a representative measurement for the samples with (red dotted line) and without (solid black line) a W spacer layer. These measurements allow us to quantify the variations of $R_{AHE}$ between samples as well as the effective perpendicular magnetic anisotropy energy $K^\perp$ by examining the out-of-plane saturation field ($B_{sat}$) in combination with the $M_s$ values reported above: $K^\perp = M_s(\mu_0 M_s - B_{sat})$, as discussed in Ref. [91]. We observe a substantial difference between the two curves in Fig. 2 (c). First, $B_{sat}$ is much larger in the presence of the W spacer, which turns out to be a general trend in the presence of a spacer layer. Figure 2 (d) shows that $K^\perp$ is reduced by about 50-75% in all the samples with spacer layers compared to the reference Pt/Co samples. We relate this substantial difference to the large perpendicular anisotropy of the Pt/Co interface, which is significantly reduced by the insertion of an ultrathin spacer. Our data also show that $K^\perp$ does not correlate simply with the atomic spin-orbit coupling constants of the different spacer elements, as expected from theoretical models of the magnetocrystalline anisotropy that take into account the width of the $d$-electron bands and hybridization effects at the Co interface[92,93]. Second, we observe that the values of $R_{AHE}$, calculated as $(R_{AHE}[\mathbf{m}_z] - R_{AHE}[-\mathbf{m}_z])/2$, are about three times lower for the samples with a spacer layer, independently of the element [Fig. 2 (e)]. Given that the AHE consists of both bulk and interface contributions[38–43], these data demonstrate that the largest contribution to the AHE originates from the Pt/Co interface. We note that the larger $R_{AHE}$ of Pt/Co cannot be ascribed to a resistivity effect[43], given that the resistance of Pt/Co is lower than that of Pt/$X$/Co (see below). The Pt/Co interface may contribute to the AHE in several ways. For instance, the magnetized Pt near Co could be one source of AHE additional to the one from bulk and interfaces of Co[42]. A second reason is that the surface-intermixed Pt/Co region can have a large AHE contribution that is absent in Pt/$X$/Co layers, similar to $Pt_xCo_{1-x}$ alloys[94]. Another source of AHE is interfacial spin-orbit coupling, which is known to induce a large AHE in Pt/Co interfaces with respect to bulk Co[38,95–97]. This would also correlate with the larger PMA found in samples without spacers. Finally, the SMR could give rise to an AHE-like contribution that would be larger in the samples without spacer. However, the latter is a less likely situation since the sign of the SMR-driven AHE is negative and its magnitude is usually 2-3 orders of magnitude smaller when considering Pt/magnetic insulator systems relative to Pt/Co bilayers [64,98,99].

Figure 2 (f) reports the square (sheet) resistance ($R_{sq}$) for all the samples, calculated as $R_0 \frac{l}{w}$ with $R_0$ being the resistance measured between the two Hall arms. Again, we observe a significant difference upon insertion of the ultrathin spacers. In Pt/Co, $R_{sq}$ is around 50-52 Ω, whereas upon insertion of the spacer layer the resistance increases to about 53-58 Ω. The higher resistance of the thicker samples is ascribed to the presence of additional interfaces, which increase the diffusive scattering and hence the overall resistance. We note that Cr, Mo and V have bulk resistivity values comparable to that of Pt and Co, whereas Ti and β-phase W are significantly more resistive than either of these two elements, which ultimately correlates with the slightly higher $R_{sq}$ measured in samples with Ti and W spacers.

### B. Magnetoresistance

We measured the longitudinal resistance using a four-point geometry by rotating the sample in a static magnetic field $B_{ext}$ = 1.8 T in three orthogonal planes (Fig. 3 a-b). This field is larger than $B_{sat}$ of all the samples, which is enough to saturate **m** along the three coordinate axes and allow us to accurately quantify the magnetoresistances $\Delta R_{xy}$, $\Delta R_{zy}$ and $\Delta R_{zx}$. Figures 3 (c)-(e) summarize the normalized magnetoresistance results expressed in % [defined as $100 \times \Delta R_{xy,zy,zx}/R_0$, with $R_0 \equiv R(\mathbf{m} \parallel \mathbf{x})$] for all samples. The largest magnetoresistance appears in the *xy* and *zy* planes, reaching 0.35-0.4% for the reference samples and 0.05-0.1% for the samples with spacer. The magnetoresistance in the *zx* plane is about one order of magnitude smaller with respect to the *xy* and *zy* planes and has opposite sign compared to the anisotropic magnetoresistance of bulk Co[100]. In other words, the resistance is higher when **m** is out-of-plane, orthogonal to **j**, and lower when **m** and **j** are collinear. Overall, in all three planes the magnetoresistance is a factor of 3 – 7 lower when a spacer layer is present, showing that the Pt/Co interface plays a crucial role in determining the amplitude of the magnetoresistance, similar to the AHE discussed earlier.

The magnetoresistive behavior of HM/FM bilayers is a subject of ongoing debate. In bulk FM materials, the resistance is typically larger when **m** is collinear with **j** due to enhanced scattering of conduction electrons from the localized *d*-orbitals (*s-d* scattering), resulting in $\Delta R_{xy} \approx \Delta R_{zx} > 0$ and $\Delta R_{zy} \approx 0$ (Ref. [101]). However, recent experiments performed on ultrathin FM films in contact with HMs typically show $\Delta R_{xy} \approx \Delta R_{zy} > 0$ and $\Delta R_{zx} \approx 0$ (Refs. [14,34,102,103]). Several explanations have been proposed for this unusual magnetoresistance. One explanation relies on the so-called anisotropic interface magnetoresistance[34], which arises due to interfacial spin scattering strongly dependent on the out-of-plane component of the magnetization, manifesting as a large $\Delta R_{zy}$. Although there are alternative models of such an effect[13,35,104–106], all such models rely on the influence of interfacial spin-orbit coupling on the scattering of electrons in multilayer systems. Another explanation relies on the spin Hall magnetoresistance (SMR)[12,102]. In this scenario, a large magnetoresistance appears in $\Delta R_{zy}$ and $\Delta R_{xy}$ due to the asymmetry in the absorption and reflection of the spin current generated by the bulk spin Hall effect of the HM upon rotation of **m** in these two planes. Common to both mechanisms, this peculiar magnetoresistance behavior arises when the HM and FM are only a few nm thick. While these mechanisms are usually discussed under separate assumptions about the origin of the spin current in HM/FM bilayers, we find that it is hard to separate them in practice, especially in systems where the spin diffusion length is comparable with the effective thickness of the interfaces. Therefore, rather than attempting such a separation, we will evidence in Sect. IV-D the correlation of the magnetoresistance and SOT properties that emerges from our measurements, without any assumption on the origin of such effects.

C. **Spin-orbit torques**

We characterize the DL-SOT and FL-SOT by measuring the current-induced effective fields $b_{DL}$ and $b_{FL}$, respectively, using the harmonic Hall voltage detection method introduced in Sect. II. and described in detail in Ref. [89]. Representative measurements of the first and second harmonic Hall resistances ($R_{\omega,H}$, $R_{2\omega,H}$) of Pt/Co are shown in Fig. 4 (a). The angular dependence of $R_{\omega,H}$ is typical of the planar Hall resistance, $R_{PHE}$, given by the second term on the right hand side of Eq. 1 and is independent of $B_{ext}$. $R_{2\omega,H}$ is strongly field dependent and includes contributions from the SOTs ($R_{2\omega,H}^{DL}$, $R_{2\omega,H}^{FL}$) and the magneto-thermal effects ($R_{2\omega,H}^{\nabla T}$). We fit $R_{2\omega,H}$ by using Eq. 2 to determine the coefficients of $\cos\varphi$ and $(2\cos^3\varphi - \cos\varphi)$, which correspond to ($R_{2\omega,H}^{DL} + R_{2\omega,H}^{\nabla T}$) and $R_{2\omega,H}^{FL}$, and plot these coefficients versus $1/B_{eff}$ and $1/B_{ext}$, respectively, as shown in Fig.4 (b) and (c). The slopes of these curves correspond to $R_{AHE}b_{DL}$ and $2R_{PHE}b_{FL}$, respectively, from which we extract $b_{DL}$ and $b_{FL}$. The intercept in Fig.4 (b) gives the magneto-thermal contribution $R_{2\omega,H}^{\nabla T}$, which we find to be negligibly small in this and all other samples studied here due to the large Pt thickness, similar to our previous reports[14,89]. Surprisingly, we also find that the linear fit of $R_{2\omega,H}$ in Fig.4 (c) has a finite unexpected offset of about -4 μΩ. At this stage, we do not have an explanation for this offset and we neglect it given that this value is much smaller than the total amplitude of the raw signal shown in (a).

Figures 4 (d) and (e) show $b_{DL}$ and $b_{FL}$ for all the samples together with the Oersted field [green dashed line in (e)] estimated by considering homogeneous current flow through the layers normalized to $j = 10^{11}$ A/m$^2$. We find that both $b_{DL}$ and $b_{FL}$ are substantially modified upon insertion of a spacer layer. We first focus on the DL-SOT. $b_{DL}$ is about 2 mT/10$^{11}$ A/m$^2$ for the reference Pt/Co samples, similar to our previous measurements[107,108], and varies between 0.6-1.6 mT/10$^{11}$ A/m$^2$ for the samples with the spacer layer. The reduction of $b_{DL}$ is larger ($\geq$ 50%) in the case of the V, Mo, and W spacer layers. Considering the trend for elements of the same group with 3$d$, 4$d$, 5$d$ valence (i.e., comparing Cr, Mo, and W), we find that the reduction in $b_{DL}$ is larger for the heavier elements, as expected due to the strong dependence of spin-orbit coupling on the atomic number. This reduction can be understood by considering different scenarios, in which $b_{DL}$ arises from the SHE in Pt, the interface-generated spin currents, or a combination of both. In fact, the insertion of a spacer layer can: i) act as an additional spin-flip scattering potential for the spin-Hall-generated spin current coming from Pt, thus reducing the resulting torque, an effect that would be particularly large for the heavier elements; ii) alter the spin current transmission/reflection probabilities; iii) generate a SHE with opposite sign to Pt; iv) alter the interface-generated spin currents due to the 'new' interface formed between the spacer layer and Pt and/or Co. As the spacer thickness is between a factor 3 to 20 lower than the spin diffusion length expected of these materials, the third scenario appears unlikely. On the other hand, i), ii) and iv) can explain the observed reduction of $b_{DL}$. The scenario described in i) corresponds to the "spin memory loss" effect, namely the transfer of spin angular momentum to the lattice due to spin-flip scattering at the interface[67]. Such an effect is known to be significant for Pt/Co and W/Co interfaces and comparatively smaller for Co interfaces with 3$d$ metal layers[75,76,109–111]. First principles calculations[61,72–74] as well as generalized magnetoelectric circuit models accounting for spin-orbit coupling at interfaces[62,112,113] show that the spin memory loss significantly alters the spin currents generated in bulk layers, but also that the interface layers, even when only a few atoms thick, generate spin currents of comparable magnitude to those generated by the "bulk" spin Hall effect. Thus, in the presence of spin-orbit coupling, effects i) and iv) likely coexist, which makes it also difficult to separate them experimentally. The scenario ii) is related to the 'spin transparency' effect, which is not related to the spin-orbit coupling but rather to the electrons' band matching that determines the spin-dependent reflection/transmission coefficients at the interface between different materials. Overall, our data suggest that one or several of these scenarios are at play here and significantly alter the SOT properties of Pt/*X*/Co relative to Pt/Co.

Within the 3*d* metal series (i.e., comparing Ti, V, and Cr), we observe no clear correlation between $b_{DL}$ and the atomic number of the 3*d* elements. The largest decrease of $b_{DL}$ is observed for the V spacer, whereas smaller effects are observed for the Ti and Cr spacers. This result is consistent with the large DL-SOT, opposite to that of Pt, reported for highly resistive β-V/CoFeB films,[52] but at variance with spin pumping measurements of YIG/Cr and YIG/V films, which report a five-fold stronger spin Hall angle for Cr compared to V (Ref.[114]). In our case, however, the strong reduction of $b_{DL}$ observed for V relative to Ti and Cr does not correlate with the increase of resistivity due to the insertion of the spacer, which is minimum for V and maximum for Ti [Fig. 2 (f)]. We thus conclude that, for the 3*d* elements, the filling of the *d*-orbitals has a stronger influence on interfacial spin-dependent scattering than the atomic number.

The dependence of the FL-SOT on the spacer layer is quite different from that of the DL-SOT. For the reference Pt/Co layers we find $b_{FL}$ of ~0.1 mT/$10^{11}$ A/m$^2$, whereas for all Pt/*X*/Co layers $b_{FL}$ changes sign and has amplitude ~-0.3 mT/$10^{11}$ A/m$^2$. In the presence of a spacer layer and independent of the element, $b_{FL}$ is thus nearly equal to the expected Oersted field, showing that the net FL-SOT almost vanishes when a spacer separates Pt and Co. After subtraction of the Oersted field, the net FL-SOT for the Pt/Co reference layers is found to be ~0.5-0.6 mT/$10^{11}$ A/m$^2$, which is about four times smaller than the DL-SOT, in agreement with previous measurements of Pt/FM bilayers with relatively thick FM[89,115]. The strongly suppressed $b_{FL}$ in the presence of a spacer layer suggests that, in this system, the FL-SOT originates predominantly at the Pt/Co interface and does not necessarily correlate with the DL-SOT. It is also interesting to note that the insertion of the spacer layer effectively reduces the proximity magnetization in Pt and the FL-SOT simultaneously. However, it has been found that the magnetic proximity effect is largely irrelevant to the magnitude of the DL and FL-SOTs in heavy metal/ferromagnet bilayers[116]. Therefore, we believe that spin-orbit coupling at the Pt/Co interface is the most likely origin of the FL-SOT, rather than the proximity magnetization of Pt.

In Fig.4 (f) we plot the relative change of $b_{DL}$ and $b_{FL}$ upon insertion of a spacer layer, which summarizes the results described above. The lack of correlation between these two sets of data clearly demonstrates the presence of multiple SOT sources in the Pt/Co bilayer system.

### D. Correlations between SOTs, magnetoresistance and perpendicular anisotropy

Analyzing the magnetotransport and SOT data together reveals interesting correlations. First, we discuss the unusual SMR-like behavior of $\Delta R_{zy}$ together with the DL-SOT. In Fig. 5 (a) we plot $\Delta R_{zy}/R_0$ as a function of $b_{DL}$. The first five pentagon-shaped points correspond to the spacer layer measurements, whereas the star is the averaged data from the five reference layers. As long as the spacer layer data are considered, we observe a linear relationship between $\Delta R_{zy}/R_0$ and $b_{DL}$, indicating a common underlying mechanism contributing to both quantities. Since $b_{DL}$ and the SMR-like behavior are predominantly associated with the interface spin accumulation due to SHE or Rashba-Edelstein effects, the correlation indicates that the $\Delta R_{zy}$ magnetoresistance is, at least partially, related to this spin accumulation. However, there is a very large difference between the extrapolation of the linear fit performed for the Pt/$X$/Co data and the data point corresponding to Pt/Co. Based on the extrapolation, only ~1/3 of $\Delta R_{zy}/R_0$ can be clearly associated with the spin accumulation in Pt/Co bilayers, meaning that the remaining ~2/3 of the magnetoresistance is related to interface scattering that is irrelevant to SOT. These data show that the magnetoresistance is a complex phenomenon in ultrathin layers and that it should not be taken as a measure of the spin Hall angle or SOT efficiency in metallic bilayers.

Another interesting correlation is found between the DL-SOT and perpendicular magnetic anisotropy. Figure 5 (b) shows that $b_{DL}$ increases linearly with $K^\perp$ in all samples with a spacer layer, and that $b_{DL}$ of Pt/Co is largest, but lies outside the linear trend. Assuming that $K^\perp$ is only determined by the element in contact with Co, our data suggest that the underlying mechanism behind the perpendicular magnetic anisotropy also plays a role in the generation of the DL-SOT. Assuming that $X$ fully separates the Pt and Co layers, the $X$/Co interface would be a new source of magnetic anisotropy and DL-SOT, which would both depend on the choice of $X$. For the elements investigated here, the additional DL-SOT would subtract to the DL-SOT arising from the SHE in Pt. With the same reasoning, assuming that the DL-SOT originates from the SHE, the interface spin-orbit coupling may influence the spin mixing conductance and spin memory loss, which finally determines the torque efficiency even though the source is the same for the systems with different spacer layers.

## IV – CONCLUSIONS

In conclusion, the SOT, AHE, magnetoresistance, magnetic anisotropy, and resistivity of Pt/Co bilayers are strongly modified by the insertion of ultrathin (0.5 nm) spacer layers of Ti, V, Cr, Mo, and W, which have opposite atomic spin-orbit coupling constant relative to Pt. The insertion of a spacer layer, independent of the element, decreases the saturation magnetization by ~15%, which we mainly associate with the decrease in the proximity magnetized Pt as it is physically separated from Co. Intermixing between Co and the spacer element could also lead to the formation of a nonmagnetic or weakly magnetic surface alloy, which would further reduce the effective magnetic Co thickness. We also find that the spacer layer significantly decreases the perpendicular magnetic anisotropy of Pt/$X$/Co relative to Pt/Co, with $K^\perp$ weakly dependent on the spacer element. Similarly, we observe a substantial drop of the AHE upon insertion of any spacer layer. This indicates that the Pt/Co interface predominantly contributes to the AHE compared to bulk Co and the $X$/Co interface. The SOTs depend strongly on the spacer element, with the most apparent trend being the monotonic decrease of the damping-like SOT with increasing atomic number in elements of the same group of the periodic table, namely Cr, Mo, and W. By contrast, the field-like SOT almost vanishes upon insertion of a spacer, independently of the element, indicating that this torque predominantly originates at the Pt/Co interface. We found a linear relationship between the damping-like SOT and the SMR-like magnetoresistance $\Delta R_{zy}$ in Pt/*X*/Co, showing that the interface spin accumulation giving rise to the former plays also an important role in the latter. The Pt/Co sample without spacer is off this linear trend, which implies that the SMR alone cannot be responsible for this magnetoresistance and that a Pt/Co interface contribution should be taken into account, whose magnitude is about twice as large as the SMR contribution. Finally, we reveal that the damping-like SOT and the interfacial perpendicular magnetic anisotropy have the same dependence on the spacer layer, suggesting a common underlying mechanism for the generation and transmission of the spin current at the Pt/Co interface and interfacial spin-orbit coupling.

## ACKNOWLEDGMENTS

We acknowledge the Swiss National Science Foundation for financial support through grant No. 200020-172775 and the National Science Foundation through grant No. ECCS-6938765.

Mater. Phys. **68**, 174405 (2003).

[82] P. Boher, F. Giron, P. Houdy, P. Beauvillain, C. Chappert, and P. Veillet, J. Appl. Phys. **70**, 5507 (1991).

[83] J.K. Tripathi, A. Kanjilal, P. Rajput, A. Gupta, and T. Som, Nucl. Instruments Methods Phys. Res. B **267**, 1608 (2009).

[84] M. Sakamaki, H. Abe, R. Sumii, K. Amemiya, T. Konishi, and T. Fujikawa, ACTA Phys. Pol. A **115**, 309 (2009).

[85] T. Liu, Y. Zhang, J.W. Cai, and H.Y. Pan, Sci. Rep. **4**, 5895 (2014).

[86] B. Cui, C. Song, G.Y. Wang, Y.Y. Wang, F. Zeng, and F. Pan, J. Alloys Compd. **559**, 112 (2013).

[87] W. Skowroński, M. Cecot, J. Kanak, S. Ziętek, T. Stobiecki, L. Yao, S. Van Dijken, T. Nozaki, K. Yakushiji, and S. Yuasa, Appl. Phys. Lett. **109**, 062407 (2016).

[88] P. Gambardella, M. Blanc, L. Bürgi, K. Kuhnke, and K. Kern, Surf. Sci. **449**, 93 (2000).

[89] C.O. Avci, K. Garello, M. Gabureac, A. Ghosh, A. Fuhrer, S.F. Alvarado, and P. Gambardella, Phys. Rev. B **90**, 224427 (2014).

[90] F. Wilhelm, P. Poulopoulos, A. Scherz, H. Wende, K. Baberschke, M. Angelakeris, N.K. Flevaris, J. Goulon, and A. Rogalev, Phys. Stat. Sol. **196**, 33 (2003).

[91] K.M. Krishnan, *Fundamentals and Applications of Magnetic Materials* (Oxford University Press, Oxford, 2016).

[92] D.S. Wang, R. Wu, and A.J. Freeman, Phys. Rev. B **47**, 14932 (1993).

[93] G.H.O. Daalderop, P.J. Kelly, and M.F.H. Schuurmans, Phys. Rev. B **50**, 9989 (1994).
22

**FIGURES**

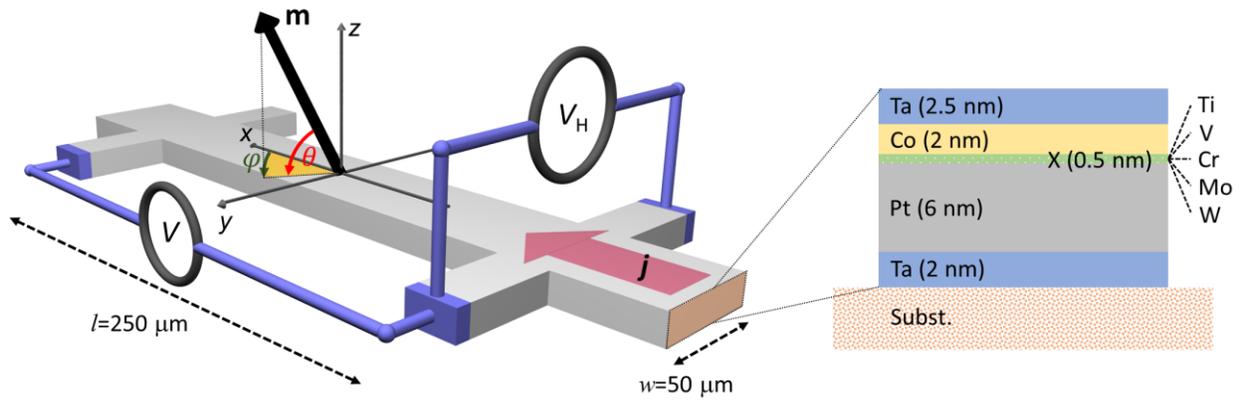

Figure 1 - (left) Device schematic, coordinate system, and electrical connections. (right) Cross-section of the sample.

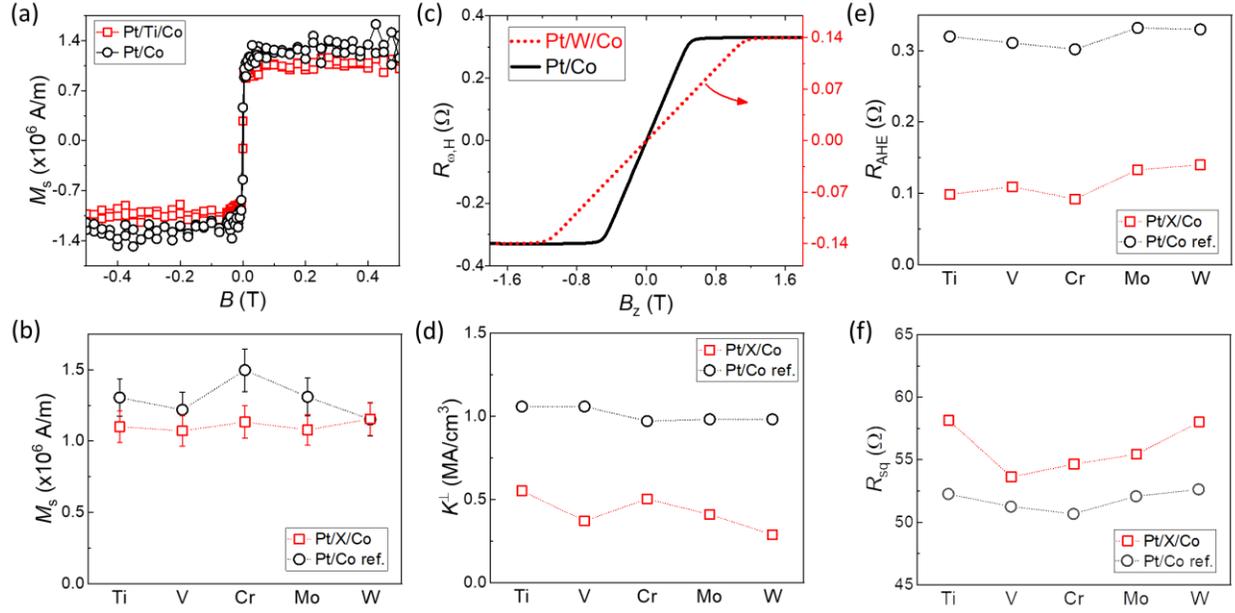

**Figure 2** - (a) Magnetization curves of Pt/Co and Pt/Ti/Co measured by vibrating sample magnetometry as a function of in-plane magnetic field. (b) $M_s$ of Pt/Co and Pt/$X$/Co. The two values for each element correspond to the measurements of the reference Pt/Co layers (black circles) co-deposited with Pt/X/Co (red squares). (c) Anomalous Hall resistance ($R_{\omega,H}$) of Pt/Co and Pt/W/Co as a function of out-of-plane field. (d) Effective perpendicular magnetic anisotropy energy ($K^\perp$) and (e) $R_{AHE}$ for all the samples extracted from measurements similar to the ones shown in (c). (f) Sheet resistance of all the samples obtained by four-point measurements.

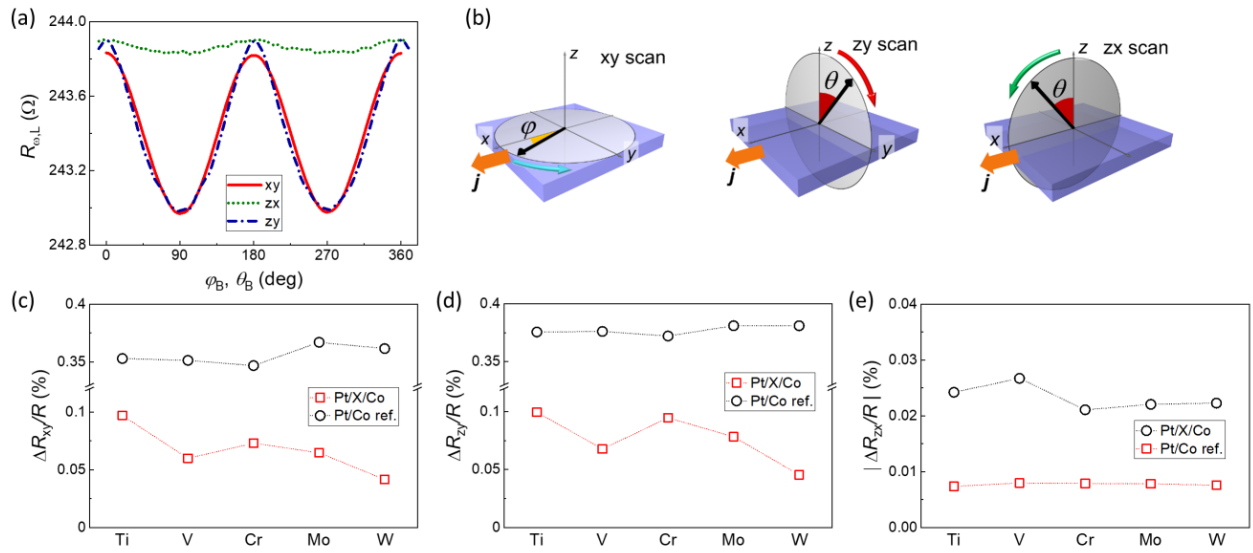

**Figure 3** – (a) First harmonic longitudinal resistance ($R_{\omega,L}$) of Pt/Co measured by rotating the sample in a fixed external field of 1.8 T. (b) Illustration of the rotation planes. (c-e) Magnetoresistance in the three planes expressed in per cent of the 240-275 $\Omega$ resistance for all the samples (note the y-axis breaks in c and d).

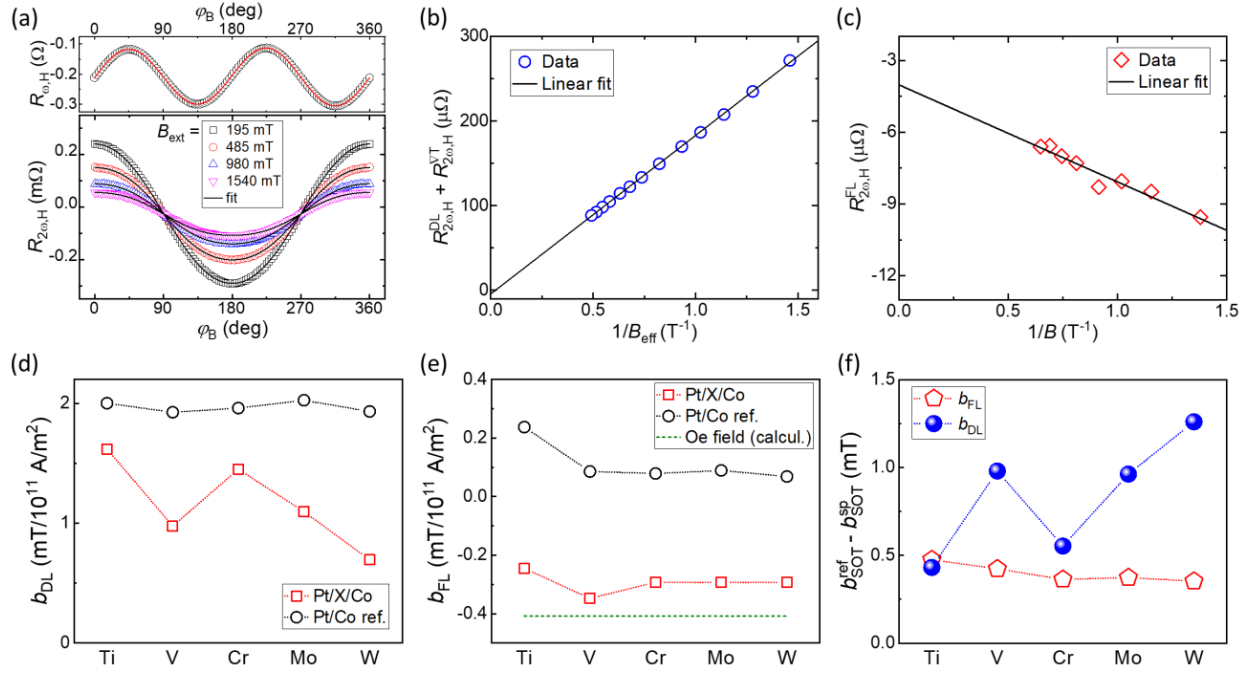

**Figure 4** – (a) First ($R_{\omega,H}$) and second ($R_{2\omega,H}$) harmonic Hall resistances of Pt/Co measured by rotating the sample in a fixed external field with various amplitudes. (b-c) Second harmonic coefficients obtained by fitting $R_{2\omega,H}$ using Eq. (1) and (2) (see text for details). (d) $b_{DL}$ and (e) $b_{FL}$ normalized to $j = 1 \times 10^7$ A/cm$^2$. (f) Difference between the values of $b_{DL}$ and $b_{FL}$ obtained in the Pt/Co and Pt/X/Co samples.

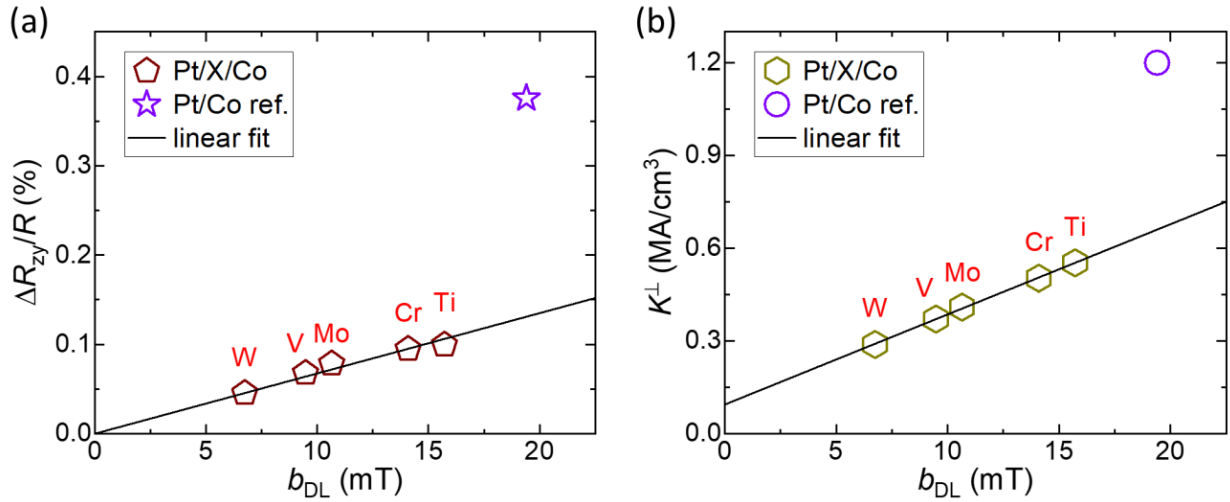

**Figure 5** – (a) Correlation between the zy-plane magnetoresistance and the damping-like SOT. The first five data points (pentagons) correspond to the Pt/$X$/Co layers; the star-shaped data point is an average of all Pt/Co bilayers. (b) Correlation between $K^\perp$ and the $b_{DL}$ obtained in Pt/Co and Pt/$X$/Co evidencing that $b_{DL}$ increases in layers with larger $K^\perp$, however, the reference Pt/Co layer does not follow the trend.